\documentclass[preprint,notoc]{JHEP3}
\usepackage{epsfig}

\def\to{\rightarrow}

\def\bi{\begin{itemize}}
\def\ei{\end{itemize}}

\def\tst{\tilde t}
\def\ttau{\tilde \tau}

\def\tz{\widetilde Z}
\def\alt{\stackrel{<}{\sim}}
\def\agt{\stackrel{>}{\sim}}
\def\be{\begin{equation}}  
\def\ee{\end{equation}}  

\title{Probing Neutralino Resonance Annihilation via \\
Indirect Detection of Dark Matter}

\author{Howard Baer and Jorge O'Farrill 
\\ Department of Physics, Florida State University Tallahassee, FL 32306, USA\\
E-mail: \email{baer@hep.fsu.edu},\email{ofarrill@hep.fsu.edu}}

\preprint{\vbox{\hbox{FSU-HEP-031130}}} 

\abstract{
The lightest neutralino of $R$-parity conserving supersymmetric 
models serves as a compelling candidate to account for the presence
of cold dark matter in the universe.
In the minimal supergravity (mSUGRA) model, 
a relic density can be found in accord
with recent WMAP data for large values of the parameter $\tan\beta$, 
where neutralino annihilation in the early universe 
occurs via the broad $s$-channel resonance of the pseudoscalar
Higgs boson $A$. We map out rates for indirect detection of
neutralinos via 1. detection of neutrinos arising from neutralino 
annihilation in the core of the earth or sun and 2. detection of gamma rays,
antiprotons and positrons arising from neutralino annihilation 
in the galactic halo. If indeed $A$-resonance annihilation is the main sink for
neutralinos in the early universe, then signals may occur in the
gamma ray, antiproton and positron channels, while a signal in the
neutrino channel would likely be absent. This is in contrast to the 
hyperbolic branch/focus point (HB/FP) region 
where {\it all} indirect detection
signals are likely to occur, and also in contrast to the stau co-annihilation
region, where {\it none} of the indirect signals are likely to occur.
}


\begin{document}


In the past decades, a growing body of astrophysical evidence 
has made an irrefragable case for the existence of cold dark matter (CDM)
in the universe\cite{review_cdm}. 
The most recent results come from the 
Wilkinson Microwave Anisotropy Probe~(WMAP)\cite{wmap}. Their results confirm 
the standard model of cosmology and fit its parameters to high precision. The 
properties of a flat universe in the $\Lambda CDM$ model are characterized by 
the density of baryons ($\Omega_b$), matter density 
($\Omega_m$), vacuum energy 
($\Omega_\Lambda$) and the expansion rate ($h$) which are measured to be:
\begin{eqnarray}
\Omega_b&=&0.044\pm 0.004 \\
\Omega_m&=&0.27\pm 0.04   \\
\Omega_\Lambda&=&0.73\pm 0.04 \\
h&=&0.71^{+0.04}_{-0.03} .
\end{eqnarray}
From the WMAP results, a value for the cold dark matter 
density of the universe can be derived:
\begin{equation}
\Omega_{CDM}h^2=0.1126^{+0.0081}_{-0.0090}
( {^{+0.0161}_{-0.0181}} )  \mbox{ at 68(95)\% CL}.
\end{equation}

A particularly attractive candidate for CDM is the lightest neutralino in
$R$-parity conserving supersymmetric models\cite{goldberg}.
In the paradigm minimal supergravity (mSUGRA) model\cite{msugra}, 
it is assumed that
at the scale $Q=M_{GUT}$, there is a common scalar mass $m_0$, 
a common gaugino mass $m_{1/2}$, and a common trilinear term $A_0$.
The soft SUSY breaking terms can be calculated at scale $Q=M_{weak}$
via renormalzation group evolution.
Electroweak symmetry breaking occurs radiatively (REWSB) due to the
large top quark mass, so that the
bilinear soft breaking term $B$ can be traded for the weak scale ratio
of Higgs vevs $\tan\beta$, and the magnitude (but not the sign) of
the superpotential $\mu$ term can be specified. Thus, the mSUGRA
model is characterized by four parameters plus a sign choice:
\be
m_0,\ m_{1/2},\ A_0,\ \tan\beta ,\ {\rm and}\ sign(\mu ) .
\ee
Once these model parameters are specified, then all sparticle
masses and mixings are determined, and scattering cross sections
may be reliably calculated.

In the early universe at very high temperatures, 
the lightest neutralino $\tz_1$ will be in thermal equilibrium, so that 
its number density is well determined. As the universe expands and cools,
there will be insufficient thermal energy to produce neutralinos, although
they can still annihilate with one another. The neutralino 
relic density can be determined by solving the Boltzmann equation for
neutralinos in a Friedmann-Robertson-Walker universe. In spite of the claims
that the lightest neutralino is a good dark matter candidate, it turns out
that in most of the parameter space of the mSUGRA model, a value of
$\Omega_{\tz_1}h^2$ well beyond the WMAP bound is generated. Only certain 
regions of the mSUGRA model parameter space give rise to a relatively low
value of $\Omega_{\tz_1}h^2$ in accord with astrophysical measurements 
and theory. These regions consist of
\footnote{Additional less prominent parameter space regions are also possible, 
such as the light higgs $h$ resonance region (bordering the 
LEP2 bounds at low $m_{1/2}$) and the top squark co-annihilation region
(for very particular choices of the $A_0$ parameter).}:
\begin{enumerate}
\item The bulk annihilation region at low values of $m_0$ and $m_{1/2}$,
where neutralino pair annihilation occurs at a large rate via $t$-channel
slepton exchange.
\item The stau co-annihilation region at low $m_0$ where 
$m_{\tz_1}\simeq m_{\ttau_1}$ so that $\tz_1$s may co-annihilate
with $\ttau_1$s in the early universe\cite{stau_co}.
\item The hyperbolic branch/focus point (HB/FP) region\cite{hb_fp} 
at large $m_0$
near the boundary of the REWSB excluded region where $|\mu |$ becomes
small, and the neutralinos have a significant higgsino component, 
which facilitates annihilations to $WW$ and $ZZ$ pairs\cite{hb_fp}.
\item The $A$-annihilation funnel, which occurs at very large 
$\tan\beta\sim 45-60$\cite{Afunnel}. 
In this case, the value of $m_A\sim 2m_{\tz_1}$.
An exact equality of the mass relation isn't necessary, since
the $A$ width can be quite large ($\Gamma_A\sim 10-50$ GeV);
then $2m_{\tz_1}$ can be several widths away from resonance,
and still achieve a large $\tz_1\tz_1\to A\to f\bar{f}$ annihilation
cross section. The heavy scalar Higgs $H$ also contributes to 
the annihilation cross section.  
\end{enumerate}

Several years ago, the bulk annihilation region of parameter space 
was favored. This situation has changed in that the bulk annihilation
region generally predicts a light SUSY Higgs boson $h$ with mass 
below LEP2 bounds, along with large- usually anomalous- predictions of
the rate for $BF(b\to s\gamma )$ decays and muon anomalous magnetic moment
$a_\mu =(g-2)_\mu/2$\cite{bbb,sug_chi2}. 
An increase of either of the parameters $m_0$
or $m_{1/2}$ leads generally to heavier sparticle masses and $m_h$
values, so that predictions for loop induced processes become
more SM-like.

Given a knowledge of which regions of model parameter space give rise
to neutralino relic densities in accord with measurements, it is
useful to examine the implications for detection of supersymmetric
matter. The direct sparticle search limits for the Fermilab
Tevatron collider\cite{tev}, the CERN LHC\cite{lhc} 
and a $\sqrt{s}=0.5-1$ TeV linear
collider\cite{nlc} have all been examined. In addition, there exist
both direct and indirect dark matter search experiments that are 
ongoing and proposed. Direct dark matter detection has been recently 
examined by many authors\cite{direct}, and observable signal rates
are generally found in either the bulk annihilation region, or in the
HB/FP region, while direct detection of DM seems unlikely in the 
$A$-funnel or in the stau co-annihilation region.

Indirect detection of neutralino dark matter\cite{eigen} may occur via
\begin{enumerate}
\item observation of high energy neutrinos originating from
$\tz_1\tz_1$ annihilations in the core of the sun or earth\cite{neut_tel}, 
\item observation of $\gamma$-rays originating from neutralino annihilation
in the galactic core or halo\cite{gamma} and 
\item observation of positrons\cite{positron} or anti-protons\cite{pbar}
originating from neutralino annihilation in the galactic halo. 
\end{enumerate}
The latter
signals would typically be non-directional due to the influence of galactic
magnetic fields, unless the neutralino annihilations occur relatively
close to earth in regions of clumpy dark matter.

The indirect signals for SUSY dark matter have been investigated 
in a large number of papers, and computer codes which yield
the various signal rates are available\cite{neutdriver,darksusy}. 
Recent works find that the various indirect signals occur at 
large rates in the now disfavored bulk annihilation region, 
and also in the HB/FP region\cite{fmw}.
Naively, this is not surprising since the same regions of parameter space
that include large neutralino annihilation cross sections in the early
universe should give large annihilation cross sections as sources of
indirect signals for SUSY dark matter. 

In this paper, we pay special attention to indirect signals for
SUSY dark matter in the $A$-annihilation funnel. We generate 
sparticle mass spectra using Isajet v7.69\cite{isajet}, which includes 
full one-loop radiative corrections
to all sparticle masses and Yukawa couplings, 
and minimizes the scalar potential using the
renormalization group improved 1-loop effective potential 
including all tadpole contributions, evaluated at an 
optimized scale choice which accounts for leading two loop terms.
Good agreement between $m_h$ values is found in comparison with
the FeynHiggs program, and there is good agreement as well in the 
$m_A$ calculation between Isajet and SoftSUSY, Spheno and Suspect codes,
as detailed in Ref. \cite{kraml}. To evaluate the indirect signals expected
from the mSUGRA model, we adopt the DarkSUSY package\cite{darksusy} interfaced
to Isasugra\footnote{Isasugra is a subprogram of the Isajet package that 
calculates sparticle mass spectra and branching fractions for a variety of
supersymmetric models}.
For our calculation of the neutralino relic density, we use the
Isared program\cite{bbb_rd} interfaced with Isajet. 
Isared calculates all relevant neutralino pair annihilation and co-annihilation
processes with relativistic thermal averaging\cite{gg}.
An important element of the calculation is that Isared 
calculates the neutralino
relic density using the Isajet 2-loop $t$, $b$ and $\tau$ Yukawa couplings
evaluated at the scale $Q=m_A$. 
The Yukawa coupling 
calculation begins with the $\overline{DR}'$ fermion masses at
scale $Q=M_Z$, and evolves via SM renormalization group equations (RGEs) to
the scale $Q_{SUSY}=\sqrt{m_{\tst_L} m_{\tst_R}}$, 
where complete MSSM 1-loop threshold
corrections are implemented. Evolution at higher mass scales is
implemented via 2-loop MSSM RGEs. The final RGE solution is gained after
iterative running of couplings and soft terms 
between $M_Z$ and $M_{GUT}$ and back until
a convergent solution is achieved.

We first show in Fig. \ref{fig1} the location of the $A$-pole in the 
mSUGRA model by plotting $|m_A-2m_{\tz_1}|/\Gamma_A$ versus $m_0$ for
$m_{1/2}=500$, 750, 1000 and 1250 GeV, for $A_0=0$ and 
{\it a}) $\tan\beta =45$ and $\mu <0$ and {\it b}) $\tan\beta =54$ and
$\mu >0$. The point in $m_0$ where $|m_A-2m_{\tz_1}|/\Gamma_A$ drops
to zero shows the location of the $A$-annihilation funnel. One may also 
note from Fig. \ref{fig1} that there is a band of $m_0$ values 
wherein $2m_{\tz_1}$ is within several widths of the $A$-pole. As 
$\tan\beta$ increases, the $b$ Yukawa coupling also increases, leading
to large values of the $A$ width $\Gamma_A$. For very large 
values of $\tan\beta\sim 54$, $\Gamma_A$ can exceed 50 GeV. 
In frame {\it a}), it can be seen that the $A$-annihilation funnel
occurs for all values of $m_{1/2}$ shown. However, in frame {\it b}),
it can be seen that the $A$-annihilation resonance is only found
for large values of $m_{1/2}\agt 1$ TeV. Even so, for lower $m_{1/2}$ 
values, the effect of the $A$-annihilation funnel is felt at the low
$m_0$ range since one may be only one to several widths away from
resonance. The explicit location of the $A$-annihilation funnel in the
$m_0\ vs.\ m_{1/2}$ plane as calculated by Isasugra is shown
in Ref. \cite{sug_chi2}, and will not be repeated here. 
\FIGURE[h]{
\epsfig{file=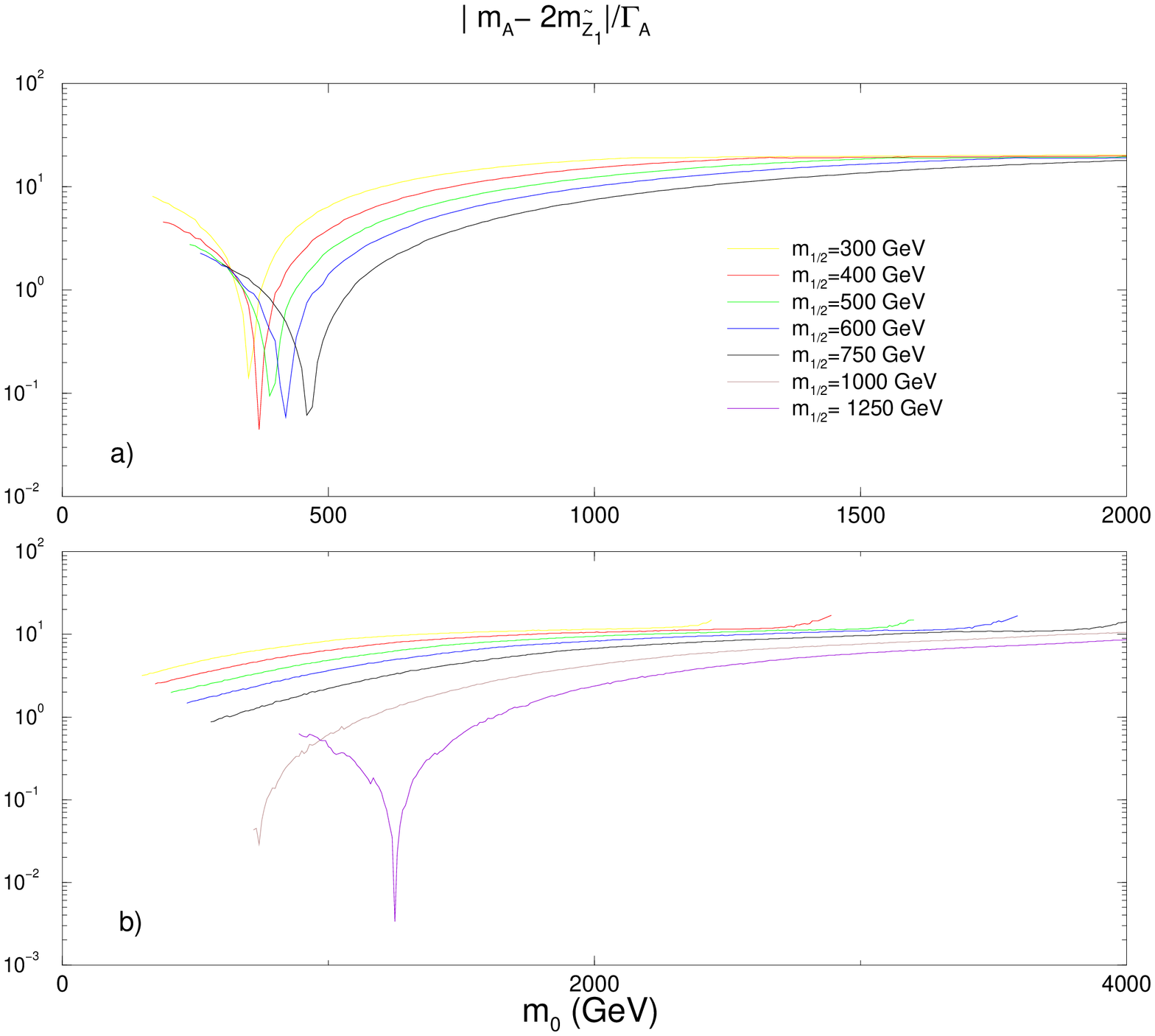,height=15cm,width=16cm,angle=-90} 
 \caption{\label{fig1}
A plot of $|m_A-2m_{\tz_1}|/\Gamma_A$ versus $m_0$ for various 
$m_{1/2}$ values and $A_0=0$ and {\it a}) $\tan\beta =45$ with
$\mu <0$ and {\it b}) $\tan\beta =54$ and $\mu >0$.}}

In Fig. \ref{fig2}, we show in the top frames the neutralino relic density
$\Omega_{\tz_1}h^2$ versus mSUGRA parameter $m_0$ for various $m_{1/2}$ values.
We also take $A_0=0$ and {\it a}) $\tan\beta =45$ and $\mu <0$ and
{\it b}) $\tan\beta =54$ and $\mu >0$. In frame {\it a}), we see that most
of the range of $m_0$ yields a value of $\Omega_{\tz_1}h^2$ far beyond the
WMAP measured range (shown by horizontal dotted lines). 
However, as $m_0$ becomes very large, one enters the 
HB/FP region, and $|\mu |$ becomes small. The $\tz_1$ gains a substantial
higgsino component which facilitates $\tz_1\tz_1$ annihilation into
final states such as $WW$, $ZZ$ and $Zh$. Thus, $\Omega_{\tz_1}h^2$ drops
into and below the WMAP measured range for CDM.
As $m_0$ decreases to small values, there is also a decrease in neutralino 
relic density. This time the decrease is due to the fact that $m_A$ 
decreases towards 
the value $2m_{\tz_1}$, so that $\tz_1\tz_1\to A,\ H\to b\bar{b}$
is enhanced. 
For low $m_{1/2}$ values, $\Omega_{\tz_1}h^2$ actually drops to very low
values--- below 0.025--- which wouldn't even be enough to
explain galactic rotation curves.
However, the $A$-annihilation funnel is quite
broad because the $b$ and $\tau$ Yukawa couplings grow with
$\tan\beta$, which increases the $A$ and $H$ widths, and also the
$\tz_1\tz_1$ annihilation rate through $s$-channel $A$ and $H$
exchange. 
Optimal values of $\Omega_{\tz_1}h^2$ in accord with the WMAP
CDM result are then achieved close to resonance, but not exactly on it.
For even lower values of $m_0$, the value of $\Omega_{\tz_1}h^2$
increases as one moves away from the $A$ resonance, until there is a final
down-turn in $\Omega_{\tz_1}h^2$ due to stau co-annihilation. 
In frame {\it b}) at the top, we show the relic density for $\mu >0$ and
$\tan\beta =54$. Again, much of the range in $m_0$ is excluded
since $\Omega_{\tz_1}h^2$ is beyond the WMAP limit. For large $m_0$,
the relic density again drops as a HB/FP region is approached. For 
low values of $m_0$, the relic density again decreases due to
the growing importance of neutralino annihilation via $s$-channel
$A$ and $H$ exchange. However, only for $m_{1/2}\agt 1$ TeV do we actually
meet the $A$ resonance. Nonetheless, for lower $m_{1/2}$ values,
the effect of the $A$ and $H$ pole is felt, and decreases the relic density
to sub-WMAP values until the stau co-annihilation region is hit at the 
very lowest values of $m_0$.
\FIGURE[h]{
\epsfig{file=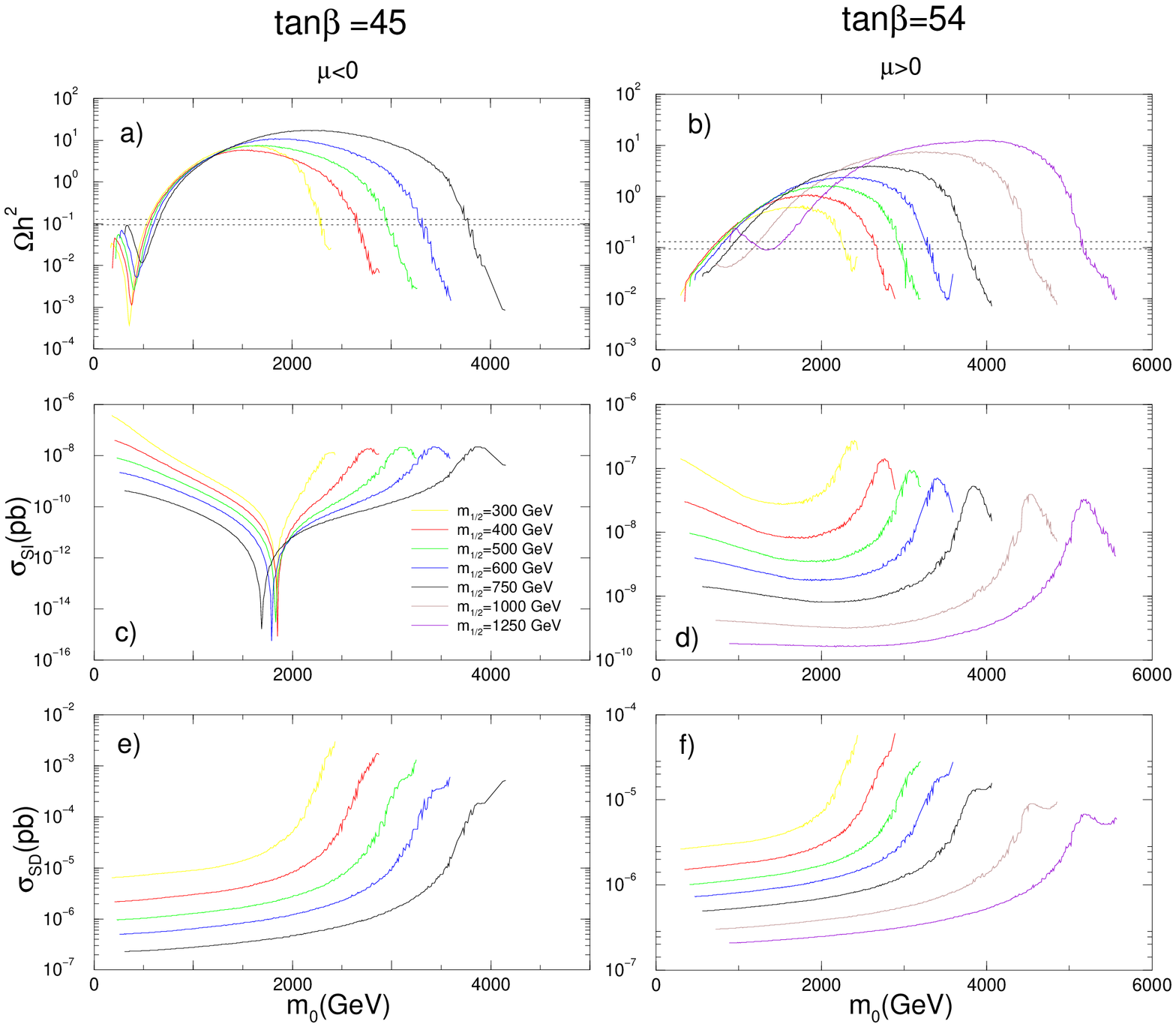,height=15cm,width=16cm,angle=-90} 
 \caption{\label{fig2}
We show the neutralino relic density 
versus $m_0$ for various values of $m_{1/2}$ and
{\it a}) $\tan\beta =45$, $\mu <0$ and {\it b}) 
$\tan\beta =53$, $\mu >0$. We take $A_0 =0$.
In frames {\it c}) and {\it d}), we show the spin-independent
neutralino-proton scattering cross section, while in frames
{\it e}) and {\it f}), we show the spin-independent neutralino-proton
scattering cross section versus $m_0$ for the same mSUGRA 
parameters as used in frames {\it a}) and {\it b}).}} 

In frames {\it c}) and {\it d}), we show the neutralino-proton
spin independent (scalar) scattering cross section $\sigma_{SI}$ 
for $\tan\beta =45$, $\mu <0$ and
$\tan\beta =54$, $\mu >0$, respectively. For low values of
$m_0$, the scattering cross section is enhanced because squark masses
become relatively light, and $u$-channel squark exchange graphs yield large
scattering amplitudes. 
For high $m_0$ values, the neutralino has a significant 
higgsino component, which enhances the $t$-channel Higgs exchange
diagrams. For the negative $\mu$ case in frame {\it c}), there is
destructive interference between Higgs and squark exchange diagrams, and
the cross section drops to zero for a particular $m_0$ value.
In frame {\it d}), there is no destructive interference, 
so the spin-independent cross section just drops to a minimal but 
finite value. We note here that Stage 3 dark matter detectors\cite{direct} 
such as Cryoarray, Zeplin-4 and Genius aim towards a sensitivity of 
roughly $10^{-9}$ pb, depending somewhat on the value of $m_{\tz_1}$.
We also note that the neutralino accretion rate for the {\it earth} depends
strongly on the spin-independent neutralino-nucleon scattering, 
which is enhanced for the heavy nuclei of which the earth is composed.

In frames {\it e}) and {\it f}), we show the neutralino-proton
spin-dependent (axial-vector) scattering cross section $\sigma_{SD}$
versus $m_0$ for the same mSUGRA parameters as the previous figures.
The cross section is again enhanced in the HB/FP region, but drops to a 
minimum as $m_0$ decreases. We note here that the neutralino accretion rate
of the {\it sun} depends strongly on the spin-dependent neutralino-nucleon
scattering cross section.

In Fig. \ref{fig3}, we show the flux of muons coming from neutralino
annihilation in the core of the earth (frames {\it a}) and {\it b})), 
and from neutralino annihilation in the core of the sun (frames
{\it c}) and {\it d})), for the same mSUGRA parameter values 
as in Fig. \ref{fig2}. Second generation neutrino telescopes such as
Icecube\cite{ice} and Antares\cite{antares} hope to probe muon flux values of
10-100 muons/km$^2$/yr\cite{eigen}. Comparing this number to the results from
frames {\it a}) and {\it b}) shows that a signal from the mSUGRA model
is unlikely to come from neutralino annihilation at the center of the earth.
The drop in muon flux at moderate $m_0$ values follows along the
curves shown previously for the neutralino-proton spin-dependent scattering
cross section. There is an enhancement in muon flux at low and high $m_0$
values, but probably not enough to create a detectable signal.
We note here that the muon flux from the earth 
shows some enhancement in rate in the $A$-annihilation funnel, but not enough
to push the expected flux levels into the observable regime.
The solid curves are plotted assuming a local relic density of 
neutralinos given by $\rho_0=0.3$ GeV/cm$^3$. 
If the relic density falls to very low values, 
then the local density may have to be rescaled in accord with
the global relic density. The dashed curves show the variation in the rates
if the local relic density is rescaled according to
$\rho=\rho_0\frac{\Omega_{\tz_1}h^2}{0.025}$ for $\Omega_{\tz_1}h^2$ values
below 0.025.
\FIGURE[h]{
\epsfig{file=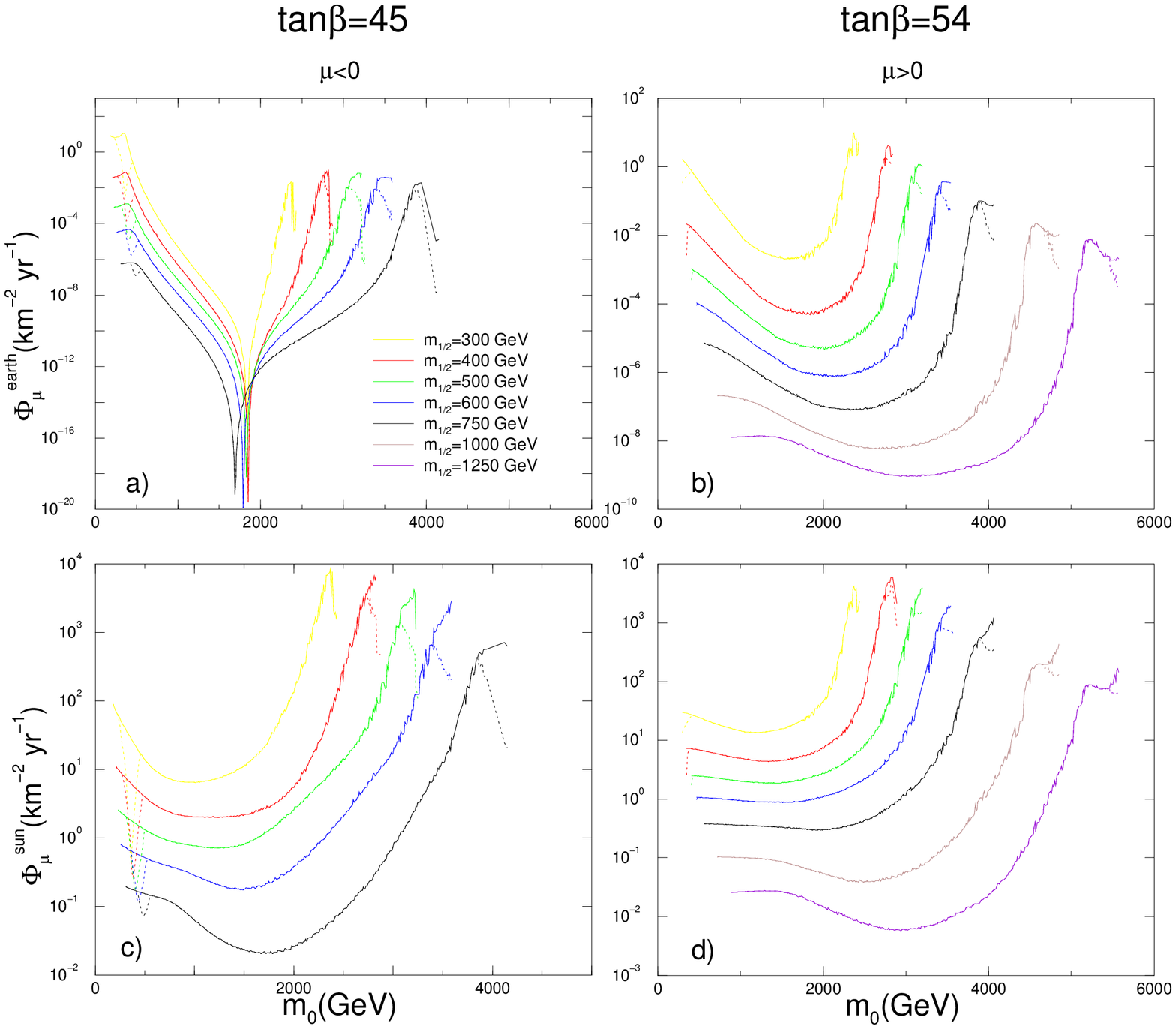,height=15cm,width=16cm,angle=-90} 
 \caption{\label{fig3}
Muon flux from the earth (frames {\it a}) and {\it b})) and sun 
(frames {\it c}) and {\it d})) 
versus $m_0$ for various values of
$m_{1/2}$ and
$\tan\beta =45$, $\mu <0$ (left-hand frames) and 
$\tan\beta =53$, $\mu >0$ (right-hand frames).
We take $A_0 =0$.
Dashed lines include rescaling of the local relic density
for values of $\Omega_{\tz_1} h^2 <0.025$.}}

Alternatively, if we examine frames {\it c}) and {\it d}), we see that
detectable levels of muon flux may indeed arise from neutralino annihilations
at the core of the sun. These large rates occur in the HB/FP region, 
as noted by other authors\cite{fmw}. It is intriguing that the large muon flux
occurs in one of the main regions where the relic density is in accord with 
WMAP analyses. However, in the $A$-annihilation funnel, there is no evidence
of enhancement of the muon flux. This is because the rate for
neutralino annihilation in the sun or earth is given by
\be
\Gamma_A ={1\over 2}C\tanh^2(\sqrt{CA}t_{\odot}),
\ee
where $C$ is the capture rate, $A$ is the total annihilation rate times
relative velocity per volume, and $t_\odot$ is the present age of the solar
system. For the sun, the age of the solar system exceeds the 
equilibration time, so $\Gamma_A\sim \frac{C}{2}$, and the muon
flux tends to follow the neutralino-nucleon scattering rate rather than the
neutralino pair annihilation cross section. The earth has typically a much
longer equilibration time, so that $\Gamma_A\sim\frac{1}{2}C^2 A t^2$,
and is hence more sensitive to the neutralino annihilation cross section
times relative velocity. 

In Fig. \ref{fig4}, we show rates for gamma rays (frames {\it a}) and
{\it b})), positrons (frames {\it c}) and {\it d}))and anti-protons
(frames {\it e}) and {\it f})) originating from neutralino annihilation 
in the galactic core and halo.
The plots are shown versus $m_0$ for the same mSUGRA parameters as
in Figs. \ref{fig2} and \ref{fig3}. In frames {\it a}) and {\it b}), 
the flux for continuum gamma rays with energy $E_\gamma >1$ GeV is shown in
units of photons/cm$^2$/sec, assuming a detector with 0.001 sr solid
angle coverage, pointed at the galactic center. 
For halo model dependence in the distribution of dark matter, we adopt
default DarkSUSY values.
Experiments such as GLAST\cite{glast}
expect to probe flux rates as low as $10^{-10}$ photons/cm$^2$/sec.
Thus, in Fig. \ref{fig4}, we see that observable rates are expected to 
occur in the HB/FP region for both $\tan\beta =45$ and $\tan\beta =54$
cases. In addition, observable rates are expected if SUSY model
parameters lie in the $A$-annihilation funnel. At low $m_0$ values,
the gamma ray flux rises and follows the $A$-annihilation resonance.
Detectable rates may occur in the $A$-funnel as long as
$m_{1/2}\alt 1$ TeV. We also show again as dashed curves the rates if
the relic density is rescaled when $\Omega_{\tz_1}h^2$ falls below
0.025. In this case, rates for gamma ray detection may fall below 
observable levels, but only because neutralinos would not be the major
constituent of CDM. If $\Omega_{\tz_1}h^2$ lies within the WMAP band, then 
typically $2m_{\tz_1}$ will lie somewhat off-resonance, and observable
rates for gamma ray detection can be found. In frame {\it b}) for
$\mu >0$, detectable rates again occur at low $m_0$ for $m_{1/2}$ values
below 500 GeV. 
\FIGURE[h]{
\epsfig{file=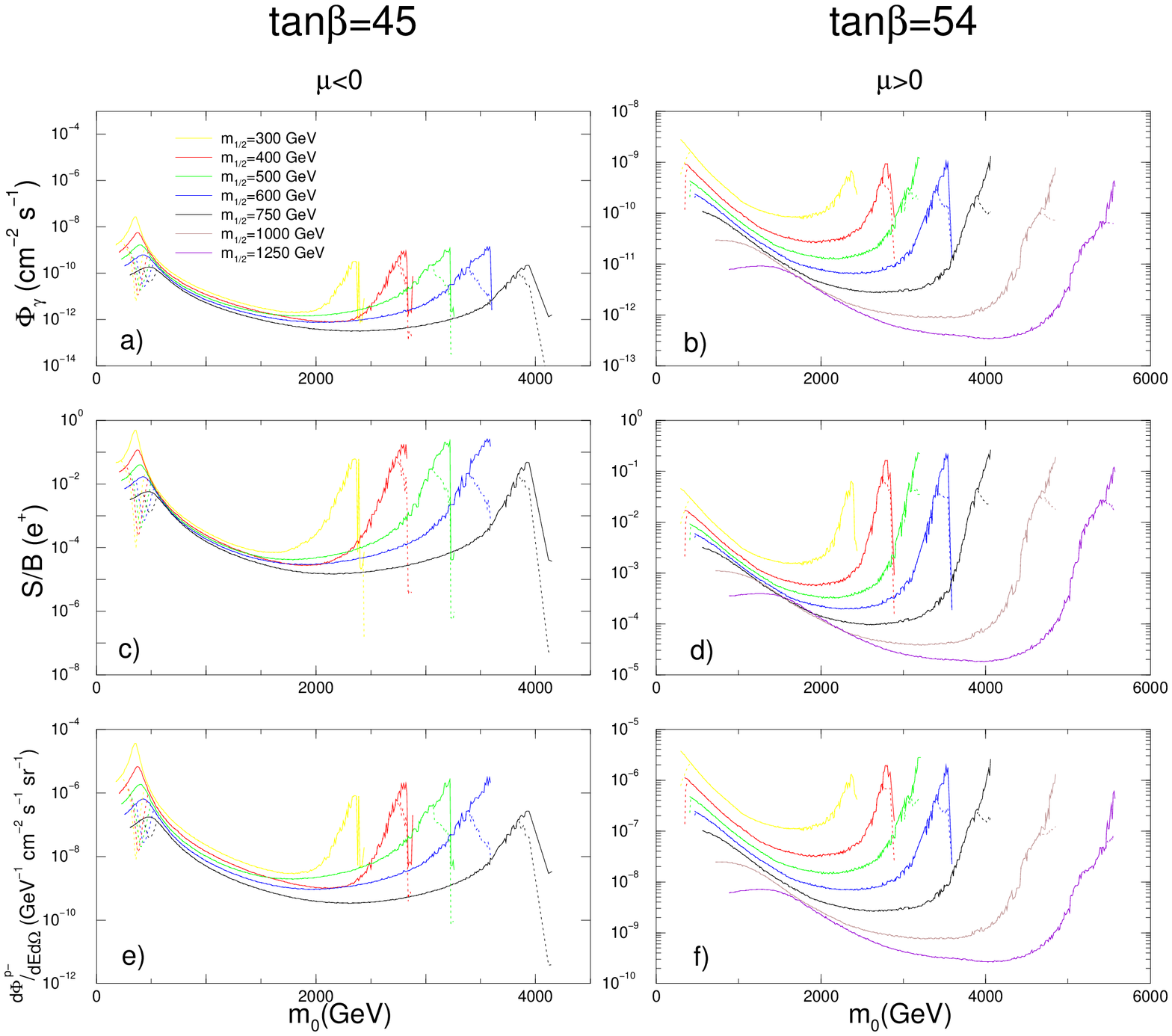,height=15cm,width=16cm,angle=-90} 
 \caption{\label{fig4}
Flux of continuum gamma rays from a $0.001 sr$ cone centered on 
the galactic center, with $E_\gamma >1$ GeV (frames {\it a}) and {\it b})).
We also show $S/B$ for positrons (frames {\it c}) and {\it d})) 
and the differential flux of antiprotons with $E_{\bar{p}}=1.76$ GeV 
(frames {\it e}) and {\it f})).
All plots are versus $m_0$ for various values of
$m_{1/2}$ and
$\tan\beta =45$, $\mu <0$ (left-hand frames) and 
$\tan\beta =53$, $\mu >0$ (right-hand frames).
We take $A_0 =0$.
Dashed lines include rescaling of the neutralino density
for values of $\Omega_{\tz_1} h^2 <0.025$.}}

In Fig. \ref{fig4} frames {\it c}) and {\it d}) we show the 
signal-to-background ($S/B$) rates for detection of positrons arising from 
neutralino annihilations in the galactic halo. To calculate the
$S/B$ rates, we adopt fit C from Ref. \cite{fmw} for the  
$E^2 d\Phi_{e^+}/d\Omega dE$ background rate. We compute the signal
using the DarkSUSY positron flux evaluated at an ``optimized'' energy
of $E=m_{\tz_1}/2$, as suggested in Ref. \cite{fmw}. A $S/B\sim 0.01$ 
rate may be detectable\cite{fmw,eigen} 
by experiments such as PAMELA\cite{pamela} 
and AMS-02\cite{ams}.
We see from Fig. \ref{fig4} that observable rates may again occur in the HB/FP
region, and also in the $A$-annihilation funnel.

Finally, in frames {\it e}) and {\it f}), we show the 
differential flux of antiprotons from
the galactic halo, $d\Phi_{\bar{p}}/dE_{\bar{p}}d\Omega$, 
for $E_{\bar{p}}=1.76$ GeV. 
In this case, background rates are more uncertain, so we
show only the differential flux of antiprotons/GeV/cm$^2$/sec/sr. 
Again, the largest rates 
occur in the HB/FP region, and also in the $A$-annihilation funnel.

We can summarize our results according to mSUGRA parameter space regions
which give rise to a reasonable relic density of CDM. In the HB/FP region,
both direct and indirect detection of SUSY dark matter is possible.
Indirect detection signals may be possible for neutrino telescopes
by observing muons arising from high energy neutrinos which are 
produced from the decays of final state originating from 
neutralino annihilation in the core of the sun. 
Detectable rates also may occur for
cosmic gamma rays, positrons and possibly anti-protons. 
If instead the $A$-annihilation funnel is the main sink for neutralinos in 
the early universe, then it is unlikely any signal would be seen from 
high energy neutrinos arising from neutralino annihilation in 
the core of the sun or the earth.
However, observable signals in the gamma ray, positron and possibly antiproton
channels may occur. Finally, if the sink for early universe neutralinos is
due to stau co-annihilation, then there may be no 
direct or indirect signals for SUSY DM. The exception occurs at very 
large $\tan\beta$ values, where the $A$-annihilation funnel and 
stau co-annihilation region begin to overlap, in which case 
gammas, positrons and antiprotons could be visible, while 
neutrino-induced muons would not be.

\section*{Acknowledgments}
 
We thank A. Belyaev, K. Matchev and X. Tata for conversations.
This research was supported in part by the U.S. Department of Energy
under contract number DE-FG02-97ER41022.
	
%

\end{document}